\documentstyle[prl,aps]{revtex}
\twocolumn
\begin{document}
\input{epsf}
\def\Fig#1{Figure \ref{#1}}
\def\Eq#1{Eq.~(\ref{#1})}
\draft
\title{Crack Front Waves and the dynamics of a rapidly moving crack}
\author{E. Sharon$^1$, G. Cohen$^2$, and J. Fineberg$^2$}
\address{$^1$The Center for Nonlinear Dynamics, The University of Texas, Austin 78712, TX, USA}
\address{$^2$The Racah Institute of Physics, The Hebrew
University of
Jerusalem,
Jerusalem 91904, Israel}
\maketitle
\begin{abstract}
Crack front waves are localized waves that propagate along the
leading edge of a crack. They are generated by the interaction of
a crack with a localized material inhomogeneity. We show that
front waves are nonlinear entities that transport energy, generate
surface structure and lead to localized velocity fluctuations.
Their existence locally imparts inertia, which is not incorporated
in current theories of fracture, to initially ``massless" cracks.
This, coupled to crack instabilities, yields both inhomogeneity
and scaling behavior within fracture surface structure.
\end{abstract}

Dynamic fracture is of fundamental and practical importance. We
consider the behavior of a crack interacting with a localized
defect. We shall show that this interaction can induce fundamental
changes to a crack's long-term dynamics. These changes imply that
a necessarily 3D view of fracture must replace the basically 2D
theory that is currently used to describe fracture in ideal
materials.

In ideal (defect-free), brittle amorphous materials, experiments
\cite{Bergqvist.74,Sharon.99,Vu} indicate that until a crack
bifurcates, its dynamic behavior is in excellent agreement with an
equation of motion \cite{Freund.90,Eshelby.70} based on a linear
elastic description of a moving crack in a 2D material. Balancing
the energy flux, $G$, per unit length of the crack with the
fracture energy, $\Gamma$, defined as the energy needed to create
a length of new fracture surface yields:
\begin{equation}\label{freundeq}
G(v,l) \sim G(l)(1-v/v_R)=\Gamma
\end{equation}
where $v$ and $v_R$ are, respectively, the instantaneous crack
velocity and the Rayleigh wave speed of the material. $G(l)$ is
dependent solely on the instantaneous crack length, $l$, and the
loading conditions. As \Eq{freundeq} has no inertial terms, a
crack tip in a 2D material should behave as a massless,
point--like object. In ideal materials, \Eq{freundeq} was shown
\cite{Sharon.99} to break down when, beyond a critical velocity,
$v_c$, a single crack loses stability
\cite{fineberg_gross1,Boudet.96,Wanner} to a state in which a
crack undergoes repetitive, short--lived microscopic branching
(``micro-branching" ) events (see e.g. \cite{Fineberg.Marder}).

\begin{figure}
\vspace{0cm} \hspace{0.0cm} \centerline{ \epsfxsize =7.5cm
\epsfysize =5.0cm \epsffile{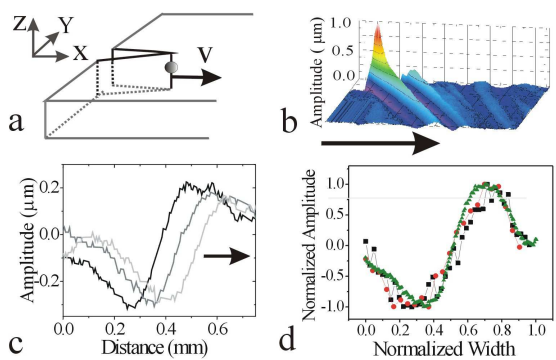} } \caption{(a) The
translational symmetry along a crack front ($z$ direction) is
broken when the front (propagating in the $x$ direction)
encounters a localized asperity. (b) The interaction with a single
asperity (located at the origin) produces localized front waves
(FW) that propagate along the front while generating tracks on the
fracture surface. Shown is a fracture surface scan of a 1.5mm x
1.5mm section of the $xz$ plane.  (c) 3 sequential profiles in the
$xy$ plane showing FW form and motion. (Time increases in
direction of arrow.) (d) FW have a unique characteristic profile.
Shown are 3 different FW whose width was normalized by the spatial
scale, $W$, of the initial asperity: $W=$ 130$\mu m$ ($\Box$)
205$\mu m$ ($\circ$) and 445$\mu m$ ($\triangle$). Arrows indicate
the propagation direction.} \label{1}
\end{figure}

Let us now consider the dynamics of a crack in a ``non—ideal"
material populated by asperities (i.e. defects which locally
perturb $\Gamma$). When a crack encounters an asperity, the
system's translational invariance normal to the propagation
direction ($z$ axis in \Fig{1}a) is broken. Thus, a crack tip can
no longer be idealized as a point—like object propagating within
an, effectively, 2D material. The dynamics of the 1D front defined
by the leading edge of the crack in a 3D material must now be
considered. In this Letter we will demonstrate that localized
waves, ``front waves", generated by an asperity, fundamentally
affect a crack's dynamics. Front waves (FW) are elastic waves that
propagate along a moving crack front. We will show that these
waves both transport energy along the crack front and locally
impart inertia to the initially ``massless" crack. This leads to a
sharp localization of the energy flux at points along the front as
well as both self--perpetuating inhomogeneities and scaling of the
fracture surface structure.
FW were first predicted as propagating velocity fluctuations
confined to the fracture plane ($y$=0 plane in \Fig{1}a).
Ramanathan and Fisher \cite{Ramanathan.97}, building on work by
Willis and Movchan \cite{Willis_Mochvan.95,Willis_Mochvan.97},
discovered that this new type of elastic wave is supported by the
linearized equations describing the perturbed stress field of a
moving crack. They are generated by asperities and propagate at
velocities, $0.94v_R< c_f < v_R$ relative to the asperity that
produced them \cite{Ramanathan.97,Morrissey.97b}. Thus, the FW
velocity, $c_{||}$, along the propagating front is $
c_{||}=\sqrt{c_f^2-v^2}$. FW are stable for $\Gamma(v)=const$ and
decay if $\Gamma(v)$ increases with $v$.  FW were also observed
numerically by Morrissey and Rice
\cite{Morrissey.97b,Morrissey.00} and, under repeated interactions
with asperities, were shown to lead to progressive roughening of
the crack front profile, in agreement with previous predictions
\cite{Perrin.94} of scalar models of fracture. Resonant effects of
FW were anticipated in \cite{boudet.00}.

Recent experiments \cite{Sharon.01} in glass (where $\Gamma(v)
\sim const$ \cite{Sharon.99}) have revealed that localized waves,
whose propagation velocity corresponds to the predicted FW, are
indeed emitted when a crack interacts with an asperity. The
observed waves have a distinct out-of-plane ($y$) component which
leaves traces along the fracture surface (\Fig{1}b). In addition,
after an initial decay, observed FW rapidly converge to
nondecaying long--lived waves with a unique characteristic profile
(\Fig{1}d). FW scale is determined by the asperity size. Their
{\em shape}, however, is independent of initial conditions. FW
retain this shape upon collisions, sustaining, like solitons, a
relative phase shift.

Our experiments were conducted in soda--lime glass plates of size
38x44x0.3cm in the $x$ (propagation), $y$ (loading) and $z$
(sample thickness) directions, respectively (see \Fig{1}a). As in
\cite{Sharon.96.2}, samples were loaded using quasistatic,  Mode I
loading. The crack velocity was measured with a resolution of 50
m/s at the plate surfaces, i.e. the $z=0$ and $z=0.3$cm $\equiv
z_{max}$ planes. Surface amplitudes were measured using a modified
Taylor-Hobson (Surtonic 3+) scanning profilometer with a 0.01$\mu
m$ resolution normal to the fracture surface. Asperities were
externally introduced within either the $z=0$  or $z_{max}$ planes
by means of scribed lines of triangular cross-section. These
lines, whose scales ranged between 100-1000 $\mu m$, locally
decreased $\Gamma$. $\Gamma$ was locally increased when these
lines were filled with Super-glue adhesive. FW were generated by
both asperity types, and above $v_c=0.42v_R$ ($v_R=3370m/s$) by
micro--branching events. Pulse profiles (e.g. \Fig{1}b,c) were
constructed in regions of constant mean velocity.

\begin{figure}
\vspace{0cm} \hspace{0.0cm} \centerline{ \epsfxsize =8.5cm
\epsfysize =2.3cm \epsffile{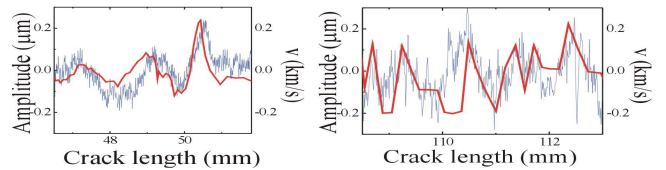} }
\vspace{.05cm}\caption{Comparison, on the $z=z_{max}$  plane, of
velocity fluctuations (bold line) and the fracture surface height
(thin line) generated by FW for (a) FW generated by an external
asperity at $v=1000m/s$ ($0.72v_c$) (b) FW generated by
micro--branching events $v=1500m/s$ ($1.08v_c$). The velocity
measurement bandwidth ($0.1\mu sec \sim 0.05-0.1mm$) was not
sufficient to capture the fine structure of the surface
measurements. } \label{2}
\end{figure}

As \Fig{1} indicates, the observed FW have a significant
out-of-plane component. Do they, in addition, correspond to
in-—plane velocity fluctuations as predicted in
\cite{Ramanathan.97,Morrissey.97b,Morrissey.00}? To ascertain
this, we measured the local velocity of the crack front on a plate
face ($z_{max}$ plane) at locations at which FW, launched from an
asperity on the opposite face (the $z=0$  plane), reached the
measurement plane. The fracture surface amplitude is compared to
the velocity fluctuations on this plane in \Fig{2}a. In \Fig{2}b
this comparison is performed for FW generated by micro--branching
events. In both cases, velocity fluctuations of 20-30\% of the
mean velocity correspond precisely to the arrival of the FW, as
indicated by the surface height measurements. Moreover, the two
signals are entirely {\em in phase}. These strong phase
correlations are seen whenever FW are generated by externally
induced asperities. These correlations are not trivial; a given
velocity peak could be generated by either a surface protrusion or
indentation. (A protrusion on one crack face corresponds to an
indentation on the other.)

We can estimate the normal velocity component $v_y \sim \delta y/
\delta t$, using the surface amplitude, $\delta y$,  (where
$\delta t \equiv \delta x/v$ for a pulse of spatial extent $\delta
x$). $v_y$ is less than 1\% of the velocity fluctuations, $v_x$,
in the propagation direction. Thus, the relatively small
out--of--plane surface (velocity) variations generated by FW
correspond to large local {\em in--plane} fluctuations of $v$.

These data further indicate that FW transport significant amounts
of energy along the front since, as \Eq{freundeq} indicates,
changes in $v$ directly correspond to changes in $G$. This
transport of energy, due to FW, explains how experiments are {\em
able} to measure velocity fluctuations when, generally, the
measurement plane at $z=0$  or $z_{max}$ is situated relatively
far from the micro--branch events, located within the plate's
interior, that initiated the fluctuations. Thus, the intrinsic
velocity fluctuations measured in experiments are, in essence,
front waves.

The strong correlation in both the amplitude and phase of the
in--plane velocity measurements and the out-of-plane surface
amplitudes indicates that these quantities can be regarded as two
components of the same wave, rather than as two independent
entities. These correlations suggest that significant coupling
exists between in--plane and out--of--plane stress components at
the crack front. As no {\em linear} coupling between in--plane and
out--of--plane stress field components exists
\cite{Willis_Mochvan.97}, the 3D nature of the FW is an indication
of their non-linear character. A further indication of FW
nonlinearity is their characteristic shape. This shape is
 obtained for sufficiently large perturbations
\cite{Sharon.01} whereas weak perturbations generate dispersive
waves that rapidly decay.

\begin{figure}
\vspace{0cm} \hspace{0.0cm} \centerline{ \epsfxsize =9cm
\epsfysize =3cm \epsffile{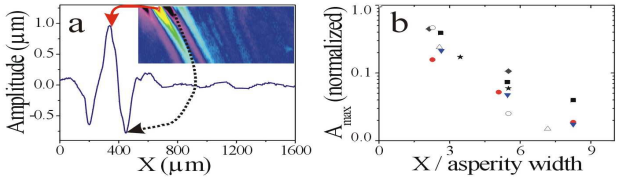} } \caption{(a) Profile (in the
$z=0$  plane) of a train of FW of decaying amplitude generated by
a single asperity. (Asperity location is between $200\mu m$ and
$400\mu m$). Note: FW are generated {\em ahead} of the initial
asperity. (inset) Topographic map in the $xz$ plane showing the
propagating FW train. The maximum amplitudes (b) of the FW
generated {\em ahead} of an asperity exponentially decay in the
propagation direction, $x$, with a characteristic decay length of
$1.8W$, where $W$ is the asperity width. Values of $W=$ $520\mu m$
($\Box$), $500\mu m$ ($\diamondsuit$) $430\mu m$ ($\nabla$)
$410\mu m$ ($\circ$) $170\mu m$ (star) $160\mu m$ (open $\Box$)
and $130\mu m$ (open $\triangle$) were used.} \label{3}
\end{figure}

In \Fig{3}  we present the surface structure generated in the
immediate vicinity of a single asperity that was induced at $z=0$
. In addition to the FW propagating away from the asperity, a
number of additional FW are typically initiated {\em directly
ahead} of the asperity. These additional waves indicate that the
crack front immediately ahead of the asperity retains a ``memory"
of the asperity's existence. This suggests that the crack front
experiences inertial effects. The highly correlated FW within
these wave--trains may explain the strong phase correlations
between velocity fluctuations and surface amplitudes evident in
\Fig{2}.

The existence of these additional waves is surprising in light of
the predictions of 2D fracture mechanics. As \Eq{freundeq}
indicates, a local change in the fracture energy should cause an
immediate corresponding change in the instantaneous velocity of a
crack. In a quasi--2D material, a crack tip can not be influenced
by the stress field fluctuations generated by its interaction with
an asperity. This is because stress variations radiate away
\cite{Vu} at the shear wave speed, $c_s$, which is considerably
larger than $v$. Therefore, the moment that a crack's tip passes
an asperity's immediate vicinity, it should instantly revert back
to its initial velocity. The wave trains in \Fig{3} indicate that
this does not occur. The exponentially decaying amplitudes of the
FW within these trains are generated {\em ahead} of the asperity.
These FW have a well--defined period which, like their decay
length ($1.8W$) and characteristic size ($\sim W$) scale ($2.3W$)
with the asperity's initial width, $W$.

\begin{figure}
\vspace{0cm} \hspace{0.0cm} \centerline{\epsfxsize =7.5cm
\epsfysize =4.8cm \epsffile{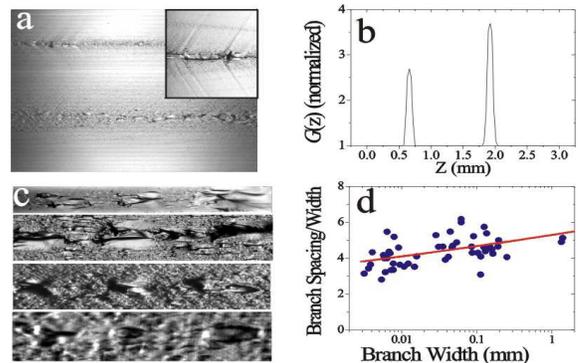} }  \caption{(a) Photograph
(of width 4mm) of two parallel branch--lines. (inset) FW emitted
from a branch--line. The energy flux ($G(z)$) distribution (b) is
highly non-uniform once branch--lines are formed. $G(z)$ was
estimated from the fracture surface presented in (a) using
measurements of micro--branch length vs micro--branch width
\protect\cite{Sharon.98}. Branch--lines are formed (c) by periodic
arrays of micro--branches at many scales. Respective photograph
heights are, from top to bottom, 2.4, 0.41, 0.12, and 0.03 mm. The
period (d) roughly scales with the branch--line width. The
predicted dependence of the branch--line period/width ratio (line)
assumes that the scale in $x$ is determined by the propagation
time of a FW across the branch--line width. Crack propagation in
(a) and (c) was from left to right.}\label{4}
\end{figure}

What is the origin of the periodicity of the FW train?  This
nontrivial behavior of the crack front ahead of the asperity
suggests that by breaking the system's translational invariance
along the crack front ($z$ direction), the initially massless
crack acquires inertia. A protrusion (indentation), breaking the
front's translational invariance, can influence (be influenced by)
other parts of the front via stress waves. As shown in
\cite{Rice.85} (for a static crack front), the local deviation,
$\delta K(z)$,  of the stress field intensity from that of a
straight front is proportional to:
\begin{equation}\label{Riceeq}
\delta K(z)\propto \int_{-\infty}^{+\infty}
(a(z')-a(z))/(z'-z)^2dz'
\end{equation}
where $(a(z')-a(z))$ is the front's deviation in the $x$ direction
at point $z$ along the front. The local value of $K(z)$ (the
``stress intensity factor")
 is proportional to $G(z)^{1/2}$ and
thereby (by \Eq{freundeq}) drives the local front velocity. Thus,
any part of the front that lags (overtakes) another will
experience an increased (decreased) local stress that tends to
stabilize a straight front. In the dynamic case, delayed
potentials will introduce inertial effects ($a(z)\Rightarrow
a(z,t)$ and $K(z)\Rightarrow K(z,t)$) thereby giving rise to
oscillatory behavior of the local stress field
\cite{Morrissey.00}. Numerical evidence for a local increase of
the stress field was provided in \cite{Morrissey.97b}, where
 a local increase of the front velocity was observed directly ahead of
an asperity. The characteristic time scale
 for stress oscillations is \cite{Morrissey.00} the time,
$W/c_{||}$, in which a front wave, travelling along the front with
velocity $c_{||}=\sqrt{c_f^2-v^2}$, traverses an asperity of size
$W$. Thus, the front immediately ahead of the asperity feels the
effects of the oscillating stress field at a spatial scale of
$Wv/c_{||}=Wv/\sqrt{c_f^2-v^2}$. This scale is consistent with the
observed scaling of both the decay length (\Fig{3}b) and spatial
periodicity (\Fig{3}a and \Fig{4}c) of the structure formed ahead
of an asperity.
 Below the micro-branching instability ($v<v_c$) the surface
structure described in \Fig{3} is typical of that generated by the
interaction of a crack front with an asperity. Above $v_c$, as
shown in \Fig{4}, micro--branching events \cite{Sharon.01}
themselves serve as FW sources since, like an asperity, they serve
to increase the local value of the fracture energy.
Micro--branches in glass, however, are {\em not} randomly
dispersed throughout the fracture surface but, as shown in
\Fig{4}a, are aligned along straight lines in the propagation
direction. Their internal structure (\Fig{4}c,d) indicates rough
periodicity in $x$ \cite{beauchamp}. As a consequence of the
increased surface area created by the micro--branches
\cite{Sharon.96.2} upon branch--line formation, the energy flux,
$G$, into the front is not evenly distributed along the front. As
\Fig{4}b shows, the total energy dissipated by a branch--line at a
given location, $z$, along the front can be significantly larger
than in the surrounding, featureless surface. This inhomogeneous
distribution of $G$, which is perpetuated for the life of a
branch--line, indicates a nonlinear focussing of energy in the $z$
direction that is not inherent in current theories of fracture. As
shown in \Fig{4}a, multiple ``branch--lines" can coexist, although
they have a tendency  to coalesce. These lines
\cite{Sharon.96.2,beauchamp,obsidian} are either initiated
spontaneously or can be triggered by an asperity.

Let us consider the behavior of a crack front immediately after a
micro-branch event/asperity of width $W$ occurs. At the conclusion
of this event the local velocity of the crack front will
momentarily ``overshoot" its unperturbed velocity $v$. For
$v<v_c$, the overshoot will exponentially ``ring down" as in
\Fig{3}. For $v>v_c$ the velocity overshoot will not decay, but
will instead initiate another branching event, directly ahead of
the first micro--branch/asperity. This scenario will again repeat
itself, thereby generating yet another branching event. In this
picture, columns of branches spaced $W v/\sqrt{c_f^2-v^2}$ apart,
in the propagation direction, will be generated. This scaling
behavior is shown in \Fig{4}c. The initial scale, $W$, of the
branch--lines is dynamically determined by $G$, as demonstrated in
\cite{Sharon.96.2,Sharon.98} where $W$ is a roughly exponential
function of $v$. As \Fig{4}d indicates, this scaling behavior is
indeed observed for over 3 orders of magnitude in $W$. The
systematic increase with $v$ of micro-branch periodicity, apparent
in the figure, is consistent with the predicted
$v/\sqrt{c_f^2-v^2}$ behavior. Note that when $v \sim v_c$ the
above picture predicts that an asperity should initiate a
branch--line. Slowly decaying branch--lines initiated by an
asperity are indeed observed in \cite{obsidian} near $v_c$

In conclusion, we have demonstrated that FW both transport energy
along a crack front and consist of coupled in--plane and
out--of--plane components. The broken translational symmetry along
the front gives rise to local inertia of the front. This local
inertia, when coupled to the micro--branching instability,
provides a mechanism for the generation of stable, directed lines
of spatially periodic micro--branches in the propagation
direction. This picture, which yields an explanation of both
branch--line periodicity and scaling, may provide insight on the
dynamic origins of fracture surface roughness \cite{Bouchaud.93}.
Both the effective focussing, by branch--lines, of energy within
the crack front and the spontaneous birth of local inertia within
a crack front point to fundamental features of crack dynamics that
cannot be incorporated in 2D descriptions of fracture.

\end{document}